\documentclass[11pt]{article}
\usepackage[margin=1in]{geometry}
\usepackage{graphicx}  %
\usepackage{amsmath}
\usepackage{slashed}
\usepackage{bm}  %
\usepackage{ulem}
\usepackage{mwe}
\usepackage{verbatim}
\usepackage[titletoc,title]{appendix}
\usepackage{comment}
\usepackage{wrapfig}
\usepackage{lipsum}

\def\Dsl{\hbox{/\kern-.6000em D}} 

\def\dsl{\,\raise.15ex\hbox{/}\mkern-13.5mu D}

\def\psip#1{\psi_{\mathbf{#1}}}

\def\ltap{\ \raise.3ex\hbox{$<$\kern-.75em\lower1ex\hbox{$\sim$}}\ }
\def\gtap{\ \raise.3ex\hbox{$>$\kern-.75em\lower1ex\hbox{$\sim$}}\ }

\def\Dslash{D\!\!\!\!\slash}

\def\dslash{\partial\!\!\!\slash}

\def\OMIT#1{}
\newcommand{\nn}{\nonumber}

\newcommand{\bea}{\begin{eqnarray}}
\newcommand{\eea}{\end{eqnarray}}

\def\mqo2{{\!\!\!}}

\def\beq{\begin{equation}}
\def\eeq#1{\label{#1}\end{equation}}
\def\eeqn{\end{equation}}

\def\beqa{\begin{eqnarray}}
\def\eeqa#1{\label{#1}\end{eqnarray}}
\def\eeqan{\end{eqnarray}}

\let\bar=\overbar

\def\Dslash{\not{\hbox{\kern-4pt $D$}}}
\def\dslash{\not{\hbox{\kern-2pt $\del$}}}

\def\msb{{\bar{\ssstyle M \kern -1pt S}}}

\def\Title#1{\begin{center} {\Large {\bf #1} } \end{center}}
\def\Author#1{\begin{center} {\normalsize {\sc #1} } \end{center}}
\def\Institution#1{\begin{center} {\normalsize {\it #1} } \end{center}}
\def\Abstract#1{\noindent {\normalsize {\bf Abstract:} {\normalfont #1}}}
\def\Conference{\vspace{4mm}\begin{raggedright} {\normalsize {\it Talk presented at the 2019 Meeting of the Division of Particles and Fields of the American Physical Society (DPF2019), July 29--August 2, 2019, Northeastern University, Boston, C1907293.} } \end{raggedright}\vspace{4mm}}

\begin{document}

%
%

\Title{Doubly Heavy Baryons and Corrections to Heavy Quark-Diquark Symmetry Prediction for Hyperfine Splitting}

\Author{ Thomas Mehen $\&$ Abhishek Mohapatra }

\Institution{Department of Physics\\ Duke University, 120 Science Drive, Durham, NC-27705, USA}

\Abstract{ In the $m_Q\rightarrow\infty$ limit, the hyperfine splittings in the ground state doubly heavy baryons $\left(QQq\right)$ and single heavy antimesons $(\bar{Q}q)$ are related by heavy quark-diquark symmetry (HQDQ) as the light degrees of freedom in both the hadrons are expected to have identical configurations. In this article, working within the framework of nonrelativistic QCD (NRQCD), we study the perturbative and nonperturbative corrections to the HQDQ symmetry hyperfine splitting relation that scale as ${\cal O}\left(\alpha_s^2\right)$ and $\Lambda_{\rm QCD}^2/m_Q^2$ respectively. In the extreme heavy quark limit, the perturbative corrections to hyperfine splitting of doubly charm or bottom baryons are a few percent or smaller. The nonperturbative corrections to hyperfine splitting are of order $10\%$ in the case of doubly charm baryons and $1\%$ or smaller in  doubly bottom baryons. }

\Conference

%
%

\section{Introduction}

The LHCb collaboration has recently reported the first observation of the doubly charm baryon $\Xi_{cc}^{++}$ with mass around $3621$ MeV \cite{Aaij:2017ueg, Aaij:2018ueg}. The doubly charm baryon was observed in two exclusive decay modes, $\Xi_{cc}^{++}\rightarrow \Lambda_c^{+}K^{-}\pi^{+}\pi^{+}$ and $\Xi_{cc}^{++}\rightarrow \Xi_c^{+}\pi^{+}$. 

The doubly heavy baryon $\left(QQq\right)$ is a bound state of two heavy quarks, $Q$, and a light quark, $q$. Thus, the appropriate theory for studying this system is nonrelativistic quantum chromodynamics (NRQCD) \cite{Bodwin:1994jh, Luke:1999kz, Brambilla:1999}. An interesting idea regarding the physics of doubly heavy baryons is that of heavy quark-diquark symmetry (HQDQ) \cite{Mehen:2006, Brambilla:2005, Savage:1990di}. In the extreme heavy quark mass limit $\left(m_Q\rightarrow\infty\right)$, the ground state of two heavy quarks in doubly heavy baryons is a tightly bound spin-1 diquark in the $\bar{3}$ of color space bound by the attractive Coulomb interaction. The size of the diquark is small compared to $\Lambda_{\rm QCD}^{-1}$ which implies that the light quark experiences the diquark as point source of color charge in $\bar{3}$ representation. This means the configuration of the light quark $q$ in the doubly heavy baryon is identical to the light quark in the heavy antimeson $\left(\bar{Q}q\right)$. This is the HQDQ symmetry which relates the properties of doubly heavy baryons and heavy antimesons in the $m_Q\rightarrow\infty$ limit.

The heavy antiquark in heavy mesons and the diquark in doubly heavy baryons have chromomagnetic couplings that are responsible for the hyperfine splittings in the ground state of these heavy hadrons. One of the implications of  HQDQ symmetry is the relation between the hyperfine splitting between spin-1/2 $\left(\Xi\right)$ and spin-3/2 $\left(\Xi^*\right)$ doubly heavy baryons and spin-0 $\left(P\right)$ and spin-1 $\left(P^*\right)$ heavy antimesons \cite{Mehen:2006, Brambilla:2005, Savage:1990di},
\begin{equation}
    m_{\Xi^*}-m_{\Xi}=\frac{3}{4}\left(m_{P^*}-m_{P}\right).
    \label{hyperfine:HQDQ}
\end{equation}
In the case of charm quarks, the recent lattice calculations predict the hyperfine splittings in doubly charm baryons to be $95$ MeV \cite{Padmanath:2015jea, Bali:2015}, whereas the HQDQ symmetry prediction from Eq.~\eqref{hyperfine:HQDQ} is around $106$ MeV.

In this proceeding, we study the perturbative and nonperturbative corrections to HQDQ symmetry prediction for the hyperfine splittings in Eq.~\eqref{hyperfine:HQDQ}. The perturbative correction arises only from the corrections to the chromomagnetic coupling of the diquark. They scale as ${\cal O}\left(\alpha_s^2\right)$ and were anticipated in Ref.\cite{Brambilla:2005}.  The nonperturbative corrections scale as $\Lambda_{\rm QCD}^2/m_Q^2$ and leads to corrections to hyperfine splittings of both heavy antimesons and doubly heavy baryons.
\section{Effective Lagrangian for Composite Diquark}
\begin{figure}[t]
	\centering
	\centerline{\includegraphics[width=10 cm,clip=true]{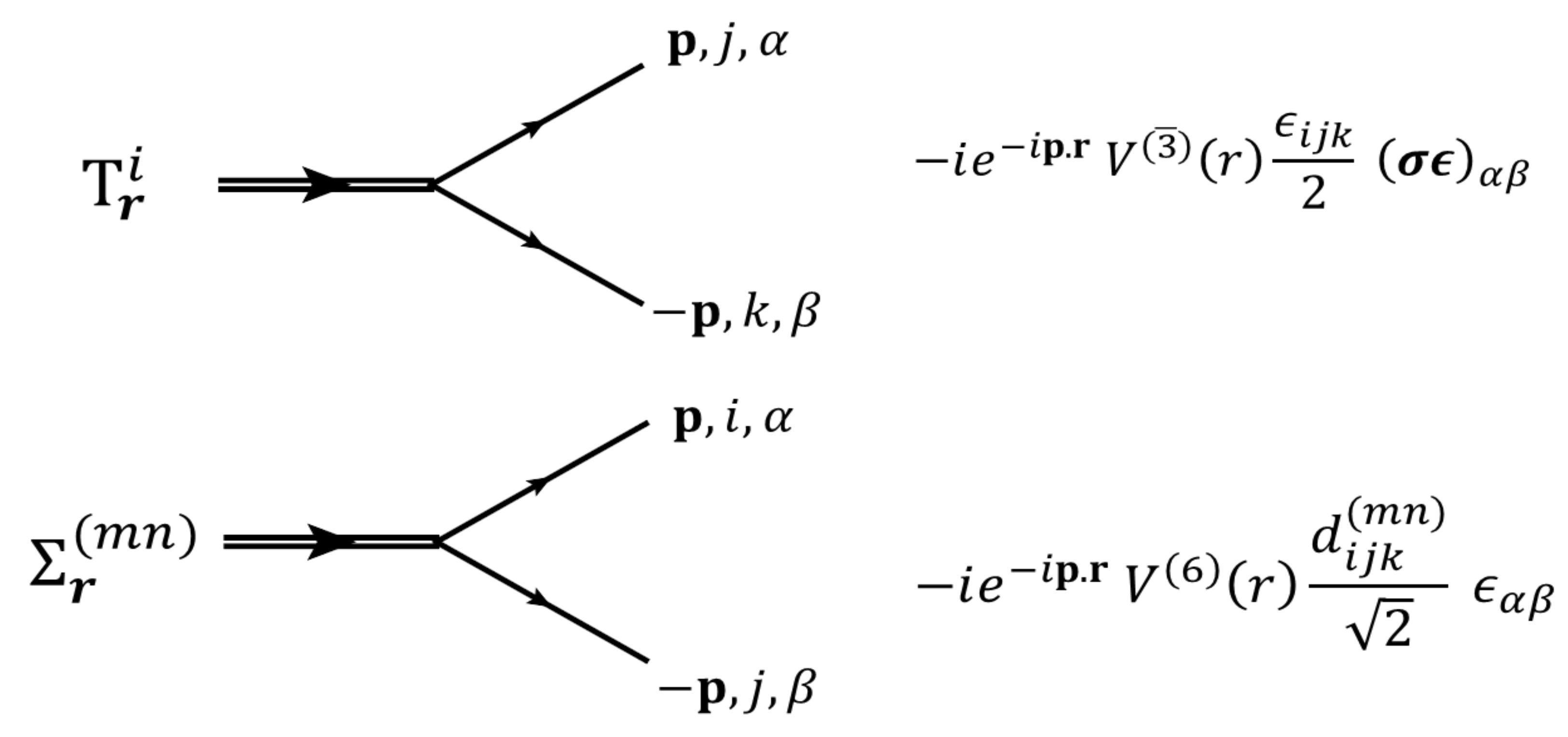} }
	\caption{The interaction vertex for coupling of the composite diquark fields to two heavy quarks.}
	\label{feynman_rules}
\end{figure}

The leading order effective Lagrangian describing the chromomagnetic coupling of diquarks that gives ${\cal O}\left(1/m_Q\right)$ corrections to heavy spin symmetry was derived in Refs.~\cite{Mehen:2006, Brambilla:2005} in the framework of NRQCD. The chromomagnetic couplings of the diquark are responsible for hyperfine splittings in the ground state of doubly heavy baryons. In this section, working within the framework of NRQCD developed in Refs.~\cite{Luke:1999kz, Mehen:2006}, we calculate the perturbative and nonperturbative corrections to the hyperfine splittings of doubly heavy baryons. 
The NRQCD Lagrangian describing the system of two heavy quarks is
\begin{eqnarray}\label{nrqcd}
{\cal L} &=&  
 -\frac{1}{4}F^{\mu\nu}F_{\mu \nu} + 
\sum_{\bf p} \psip p ^\dagger   \Biggl ( i D^0 - \frac{\left({\bf p}-i{\bf D}\right)^2}{2 m_Q} 
 + \frac{g}{2 m_Q} \,\bm{\sigma}\cdot {{\bf B}} 
 \Biggr )
 \psip p  \nonumber \\
&& - \frac{1}{2} \sum_{\bf p,q}
 \frac{g_s^2}{ \bf (p-q)^2} 
  \psip q ^\dagger  T^A \psip p \psip {-q}^\dagger  T^A \psip {-p} + \ldots \, ,
\end{eqnarray}
where $\psip p$ represents the quark field with a three vector label ${\bm p}$, ${\bm B}$ is the chromomagnetic field, and the ellipsis represents the higher order interactions as well as terms including soft gluons. Using the spin and color Fierz identities, the quartic interaction term in Eq.~\eqref{nrqcd} can be written in terms of quark billinears of definite spin and in anti-triplet $\left(\bar 3\right)$ and sextet $\left(6\right)$ color configurations. 
After performing a Fourier transform, with respect to the momentum labels and using the Hubbard–Stratonovich transformation, the quartic interaction term in Eq.~\eqref{nrqcd} can be eliminated in favor of composite diquark fields
\bea
{\cal L} &=&  
 -\frac{1}{4}F^{\mu\nu}F_{\mu \nu} + 
 \sum_{\bf p} \psip p ^\dagger   \Biggl( i D^0 - \frac{\left({\bf p}-i{\bf D}\right)^2}{2 m_Q} 
 + \frac{g}{2 m_Q} \, \bm{\sigma}\cdot {{\bf B}} 
 \Biggr)\psip p  \nn\\
 &+&\frac{1}{2} \int d^3 {\bf r} \, V^{\left(\bm{\bar{3}}\right)}(r)\Biggl(
 {\bf T}^{i\dagger}_{\bf r}{\bf T}^i_{\bf r}-{\bf T}^{i\dagger}_{\bf r}\sum_{\bf p} e^{i \bf p\cdot r} 
\frac{1}{2} \epsilon_{ilm}(\psi_{\bf -p})_l \epsilon \bm{\sigma}(\psi_{\bf p})_m \nn\\
&&\hspace{4.2 cm}- \sum_{\bf q} e^{-i \bf q\cdot r} \epsilon_{ijk}\frac{1}{2}(\psi^\dagger_{\bf q})_j  \bm{\sigma} \epsilon (\psi^\dagger_{\bf -q})_k{\bf T}^i_{\bf r}\Biggr)\nn\\
&+& \frac{1}{2} \int d^3 {\bf r} \, V^{({\bm  6})}(r)
\Biggl(\Sigma^{(mn)\dagger}_{\bf r} \Sigma^{(mn)}_{\bf r}-
\Sigma^{(mn)\dagger}_{\bf r}\sum_{\bf p} e^{i \bf p\cdot r} \frac{1}{\sqrt{2}} \, d^{(mn)}_{ij} (\psi_{\bf-p})_i \epsilon^T (\psi_{\bf p})_j\nn\\
&&\hspace{4.2 cm}-\sum_{\bf q} e^{-i \bf q \cdot r} \frac{1}{\sqrt{2}} \,
d^{(mn)}_{ij} (\psi_{\bf q})_i \epsilon (\psi_{\bf -q})_j\Sigma^{(mn)}_{\bf r}\Biggr),
\label{Lagrangian:HS}
 \eea
where we have suppressed the spin indices, the Roman letters refer to color indices, $\sigma^{i}$ is the Pauli matrix, ${\bm \epsilon}=i\sigma^2$ is an anti-symmetric 2 $\times 2$ matrix, $d_{ij}^{\left(mn\right)}$ are symmetric matrices in color space, $\epsilon_{ijk}$ are anti-symmetric matrices in color space and $r$ is the distance between the two heavy quarks in the diquark. The anti-triplet potential $V^{\left(\bm{\bar{3}}\right)}\left(r\right)$ and the sextet potential $V^{\left({\bm 6}\right)}\left(r\right)$ are given by 
\bea
V^{(\bm {\bar{3}})}(r) = -\frac{2}{3} \frac{\alpha_s}{r}, \qquad V^{({\bm 6})}(r) = \frac{1}{3} \frac{\alpha_s}{r} \, .
\eea
In Eq.~\eqref{Lagrangian:HS}, the composite diquark field $T_{\bm r}^i$ is a spin-$1$ field in the $\bar{3}$ of color and $\Sigma_{\bm r}^{mn}$ is a spin-$0$ diquark field in the $6$ color of :
\begin{align}
 T_{\bm r}^{i}&=\sum_{\bf p} e^{i \bf p \cdot r} \frac{1}{2}\, \epsilon^{ijk} (\psi_{\bf -p})_j \epsilon \, \bm{\sigma} (\psi_{\bf p})_k,\label{spin1_diquark}\\
 \Sigma^{\left(mn\right)}_{\bm r}&= \sum_{\bf p} e^{i \bf p \cdot r} \frac{1}{\sqrt{2}} \, 
d^{(mn)}_{ij} (\psi_{\bf-p})_i \epsilon^T (\psi_{\bf p})_j.
\label{spin0_diquark}
\end{align}
The Feynman rules for the interaction of composite diquark fields with two heavy quarks are shown in Fig.\ref{feynman_rules}.

\subsection{Perturbative corrections to hyperfine splittings}

The chromomagnetic coupling of heavy quarks that is responsible for the hyperfine splitting in the ground state of heavy mesons is given by the ${\bm \sigma}\cdot{\bm B}$ term in Eq.~\eqref{nrqcd}. The leading order (LO) Lagrangian for the chromomagnetic coupling of diquarks that is responsible for hyperfine splitting in ground state doubly heavy baryons was derived in Refs.~\cite{Mehen:2006, Brambilla:2005}:
\begin{equation}
{\cal L}_{\sigma.{\bm B}} = i \, \frac{g}{2 m_Q}\int d^3 {\bf r} \, {{\bf T}}^{i\,\dagger}_{\bf r} \cdot{\bf B}^c \,\bar{T}^c_{ij} \times{\bf T}^j_{\bf r} .
\label{chromo_coupling}
\end{equation}
The effective Lagrangian in above equation contributes at ${\cal O}\left(v^2\right)$ to effective action in NRQCD power counting. The composite diquark field $\Sigma_{\bm r}$ is a scalar and thus doesn't have any chromomagnetic coupling. 

\begin{figure}[h!]
	\centering
	\centerline{ \includegraphics[width=6 cm,clip=true]{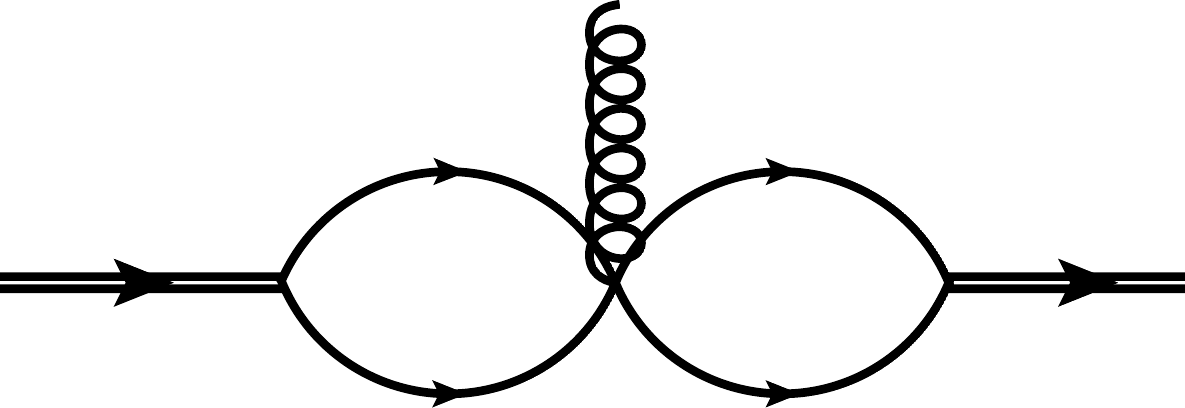} }
	\vspace*{0.0cm}
	\caption{The two-loop diagram that leads to NLO chromomagnetic coupling of diquark in Eq.~\eqref{chromo_correction}}.
	\label{two_loop}
\end{figure}

\begin{figure}[h!]
	\centering
	\centerline{ \includegraphics[width=5 cm,clip=true]{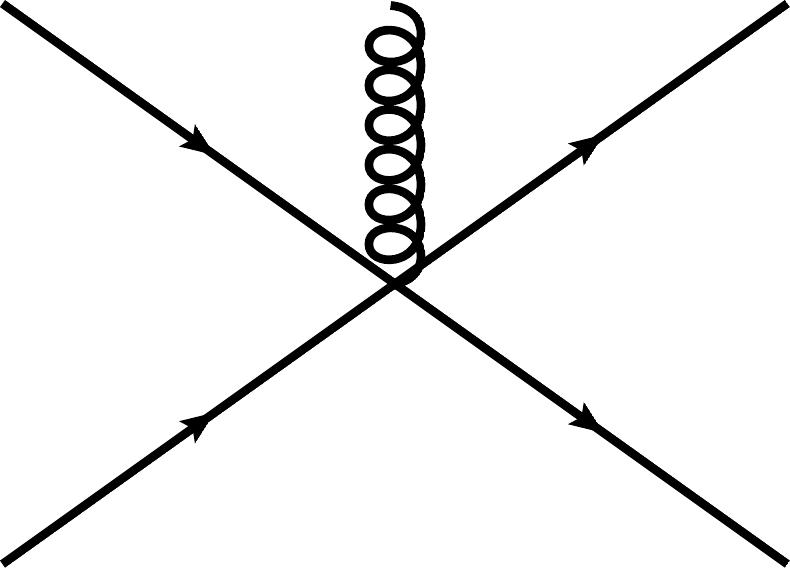} }
	\caption{Effective 5-point contact vertex in NRQCD.}
	\label{effective_operator}
\end{figure}

The perturbative corrections to the hyperfine splittings in doubly heavy baryons comes from a next-to-leading order (NLO) chromomagnetic coupling of diquarks that contributes to the effective action at ${\cal O}\left(v^4\right)$ in NRQCD power counting. The NLO coupling comes from a two-loop diagram shown in Fig.~\ref{two_loop}, which  has an effective 5-point contact interaction shown in Fig.~\ref{effective_operator}.
\newpage
This interaction comes from matching tree diagrams for low-energy  $QQ\rightarrow QQg$ scattering in full QCD onto NRQCD \cite{Mohapatra:2019}, and is given by 
\begin{equation}
    {\cal L}_{\mathrm eff}=-\frac{g^3}{2}\frac{1}{2m_Q}\sum_{{\bm P_1},{\bm P_2},{\bm P_3},{\bm P_4}}\psi^{\dagger}_{\bm P_4}T^{a}\psi_{\bm P_2}\psi^{\dagger}_{\bm P_3}\left(T^{a}T^{c}+T^{c}T^{a}\right)\frac{{\bm \sigma}\cdot{\bm B}^{c}}{2m_Q}\psi_{\bm P_1}\frac{1}{\left({\bm P_4}-{\bm P_2}\right)^2}.
    \label{effective_operator_Lagrangian}
\end{equation}

After evaluating the two-loop diagram in Fig.~\ref{two_loop}, the NLO Lagrangian describing the chromomagnetic coupling of diquark is \cite{Mohapatra:2019}
\begin{equation}
    {\cal L}^{'}_{\sigma.{\bm B}}=i \, \frac{g}{2m_Q} \,\frac{\alpha_s}{3 m_Q}\int d^3 {\bf r} \, {{\bf T}}^{i\,\dagger}_{\bf r} \cdot \frac{1}{r}\,{\bf B}^c \, \bar{T}^c_{ij} \times{\bf T}^j_{\bf r}.
    \label{chromo_correction}
\end{equation}
Naively, one would expect the above NLO Lagrangian to scale as $1/m_Q^2$ but instead it scales as $1/m_Q^2\times\alpha_s/r$. Since $r^{-1}\sim m_Q\alpha_s$ in the diquark, the NLO chromomagnetic coupling of diquark in Eq.~\eqref{chromo_correction} gives an ${\cal O}\left(\alpha_s^2\right)$ correction to hyperfine splitting.

In the absence of the chromomagnetic coupling of the heavy quark in Eq.~\eqref{nrqcd}, the spin-0 meson $D$ and spin-1 meson $D^*$ are degenerate. Similarly, the spin-1/2 baryon $\Xi$ and spin-3/2 baryon $\Xi^*$ are degenerate in the absence of the chromomagnetic coupling of diquarks in Eqs.~\eqref{chromo_coupling} and \eqref{chromo_correction}. The hyperfine splittings in heavy antimesons and doubly heavy baryons depend on the matrix elements of the chromomagnetic coupling of quarks and diquarks. The matrix element of the NLO chromomagnetic coupling of the diquark in Eq.~\eqref{chromo_correction} depends on $\langle 1/r\rangle$ which can be calculated in the extreme heavy quark mass limit in which the ground state spatial wavefunction of the diquark in the doubly heavy baryon is given by the s-wave hydrogen like wavefunction
\begin{equation}
\phi\left({\bm r}\right)=\left(\frac{1}{\pi a_0^3}\right)^{1/2}e^{-r/a_0},
\label{eq:diquark_wavefunction}
\end{equation}
 where $a_0$ is the Bohr radius 
\begin{equation}
    a_0=\frac{3}{\alpha_s m_Q},
    \label{eq:Bohr}
\end{equation}
Therefore, due to the perturbative correction from the NLO chromomagnetic coupling of the diquark in Eq.~\eqref{chromo_correction}, the HQDQ symmetry prediction of hyperfine splitting in Eq.~\eqref{hyperfine:HQDQ} is modified to
\begin{align}
    m_{D^*}-m_D&=\frac{4}{3}\left(m_{\Xi^*}-m_{\Xi}\right)\left(1+\frac{\alpha_s\left(m_Q\right)}{3m_Q}\left\langle\frac{1}{r}\right\rangle\right),\nonumber\\
    &=\frac{4}{3}\left(m_{\Xi^*}-m_{\Xi}\right)\left(1+\frac{\alpha_s\left(m_Q\right)\alpha_s\left(m_Qv\right)}{9}\right),
    \label{eq:hyperfine}
\end{align}
The two factors of coupling constant $\alpha_s$ in Eq.~\eqref{eq:hyperfine} are at two different energy scales. One of the $\alpha_s$ is due to matching QCD onto NRQCD at the energy scale $m_Q$ and the other $\alpha_s$ is obtained from the evaluation of matrix elements, for which the appropriate scale is $m_Qv$. For simplicity, the coupling constant $\alpha_s$ is evaluated at the energy scale $m_Qv$. If the doubly heavy baryon is composed of charm quark, then $\alpha_s\left(m_Qv\right)\approx0.52$, which implies that the perturbative correction to the hyperfine splitting is $\approx3.1\times10^{-2}$. If the doubly heavy baryon is composed of bottom quark, then $\alpha_s\left(m_Qv\right)\approx0.35$, which implies that the perturbative correction to the hyperfine splitting is $\approx1.4\times10^{-2}$. Therefore, we  conclude that the perturbative corrections to HQDQ symmetry prediction in Eq.~\eqref{hyperfine:HQDQ} are very small in the limit $m_Q\rightarrow\infty$.
\subsection{Nonperturbative corrections to hyperfine splitting}



The nonperturbative corrections to the hyperfine splitting scale with powers of $\Lambda_{\rm QCD}/m_Q$. To 
${\cal O}\left(\Lambda_{\rm QCD}^2/m_Q^2\right)$, the heavy quark spin symmetry violating operators in the heavy quark effective theory (HQET) Lagrangian is
\bea
\frac{g}{2 m_Q} \sum_{\bf p}\psi_{\bf p}^\dagger \, \bm{\sigma} \cdot {\bf B} \, \psi_{\bf p} + i\frac{g}{8 m_Q^2}\sum_{\bf p} \psi^\dagger_{\bf p} \, \bm{ \sigma} \cdot ({\bf D\times E - E \times D}) \, \psi_{\bf p}  \,.
\eea
The heavy quark field $\psi_{\bf p}$ in the second operator couples to a different background field: ${\bf B} \to i({\bf D\times E -   E \times D})/(4 m_Q)$. Thus, after following exactly the calculation in Ref.~\cite{Mehen:2006}, the heavy quark spin symmetry violating Lagrangian for the diquark field is given by
\begin{equation}
    {\cal L} =i \, \frac{g}{2m_Q}  \int d^3 {\bf r} \, {{\bf T}}^{i\,\dagger}_{\bf r} \cdot \,{\bf B}^c \, \bar{T}^c_{ij} \times{\bf T}^j_{\bf r} -
    \frac{g}{8m_Q^2} \int d^3 {\bf r} \, {{\bf T}}^{i\,\dagger}_{\bf r} \cdot \,({\bf D\times E}^c - {\bf D\times E}^c )\, \bar{T}^c_{ij} \times{\bf T}^j_{\bf r}.
    \label{TLagatmsquared}
\end{equation}
Since both the operators in the above Lagrangian have identical spin and color structures, the factor of $3/4$ relating two hyperfine splittings in Eq.~\eqref{hyperfine:HQDQ} remains unaffected.

There are other spin symmetry violating operators in HQET at ${\cal O}\left(1/m_Q^3\right)$ and higher but those have different spin and color structure compared to the two operators in Eq.~\eqref{TLagatmsquared}. Thus, we expect the corrections to HQDQ symmetry prediction for hyperfine splitting in Eq.~\eqref{hyperfine:HQDQ} to be ${\cal O}\left(\Lambda_{\rm QCD}^2/m_Q^2\right)$, which will be $10\%$ for doubly charm baryons and $1\%$ or smaller for doubly bottom baryons.

\section{Conclusions}
The hyperfine splittings in doubly heavy baryons $\left(QQq\right)$ and heavy antimesons $\left(\bar{Q}q\right)$ are related by HQDQ symmetry as shown in Eq.~\eqref{hyperfine:HQDQ} in the $m_Q\rightarrow\infty$ limit. In this proceeding, we study the perturbative and nonperturbative corrections to HQDQ symmetry prediction for hyperfine splitting in Eq.~\eqref{hyperfine:HQDQ}. The perturbative corrections to hyperfine splitting scale as ${\cal O}\left(\alpha_s^2\right)$ in the $m_Q\rightarrow\infty$ limit and is given by Eq.~\eqref{eq:hyperfine}. This ${\cal O}\left(\alpha_s^2\right)$ correction comes from an effective 5-point contact interaction shown in Fig.~\ref{effective_operator} and the NLO effective Lagrangian for the chromomagnetic coupling of diquark is given in Eq.~\eqref{chromo_correction}. 
The perturbative correction to  hyperfine splitting is $\approx3\times10^{-2}$ in case of doubly charm baryon $\approx1.4\times10^{-2}$ in case of doubly bottom baryons. 
We also gave an argument explaining why the nonperturbative  corrections to Eq.~(\ref{hyperfine:HQDQ})  should scale as ${\cal O}(\Lambda_{\rm QCD}^2/m_Q^2)$ rather than ${\cal O}(\Lambda_{\rm QCD}/m_Q)$ as argued in Ref.~\cite{Brambilla:2005} . Therefore, the nonperturbative corrections to hyperfine splitting relation in Eq.~(\ref{hyperfine:HQDQ}) is of order 10\% for doubly charm baryons  and 1\% or smaller  for doubly bottom baryons. The estimate for the nonperturbative corrections is consistent with the recent lattice calculations of doubly charm spectra  \cite{Padmanath:2015jea,Bali:2015}. In order to verify the scaling of nonperturbative corrections to hyperfine splittings with $m_Q$, it would be interesting to perform lattice calculations of hyperfine splittings in doubly heavy baryons in which the heavy quark mass $m_Q$ is varied. 

\section*{Acknowledgements}
This research is supported in part by  Director, Office of Science, Office of Nuclear Physics, of the U.S. Department of Energy under grant number
DE-FG02-05ER41368.

\end{document}